October 23, 2023

# The SPARC Primary Reference Discharge defined by cfsPOPCON

T. Body, C. Hasse, A. Creely

Comparing the Primary Reference Discharge (PRD) design point for SPARC as defined by the "v4.0.0" open-source release version of cfsPOPCON to the SPARC Physics Basis design point.

## Brief

The Primary Reference Discharge (PRD) is a design point maximizing the highest-achievable fusion power gain in a full-field (12.2T), full-current (8.7MA) DT-fueled H-mode on SPARC. The Primary Reference Discharge was defined in the SPARC Physics Basis [1] using a precursor to the cfsPOPCON scoping tool. In this memo, we discuss modifications made to the cfsPOPCON scoping tool since the release of the SPARC Physics Basis and show that the key parameters of the PRD design point have not changed significantly.

### The SPARC Primary Reference Discharge defined by the SPARC Physics Basis.

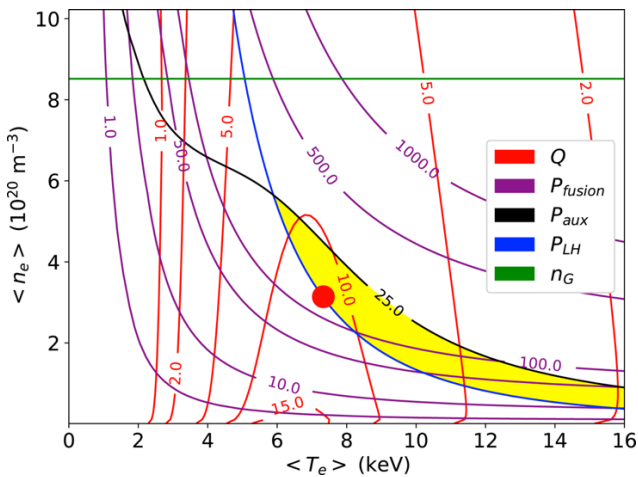

Figure 1: The PRD as defined in SPARC Physics Basis [1]. Originally figure 4 from Creely et al., 2020. Original caption: Plasma operating contour (POPCON) for full-field, full-current H-mode operation in SPARC. Red contours are Q, purple contours are fusion power in MW, the black contour is the available auxiliary heating power in MW, the blue contour is the L–H threshold power and the green contour is the Greenwald density limit. The yellow shaded region represents the operational space where SPARC is above the L–H power threshold but below the available auxiliary heating power. Temperature and density are the volume-averaged values. The red circle is the operating point for the full-performance H-mode discharge.

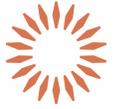

# The SPARC Primary Reference Discharge defined by the "v4.0.0" release of cfsPOPCON.

## Reproducing this result.

The analysis in this section is included as an example case in cfsPOPCON. To run this example, install cfsPOPCON and then run:

```
poetry run popcon example_cases/SPARC_PRD -p example_cases/SPARC_PRD/plot_popcon.yaml –show
```

This runs on a reduced resolution grid. To exactly reproduce the analysis below, replace the `grid` block in `example_cases/SPARC_PRD/input.yaml` to:

```yaml
grid:
  average_electron_density:
    # Average electron density in 1e19 particles / m^3
    min: 1.0
    max: 100.0
    num: 100
    spacing: linear
  average_electron_temp:
    # Average electron temperature in keV
    min: 0.5
    max: 16.0
    num: 100
    spacing: linear
```

## The result from cfsPOPCON v4.0.0.

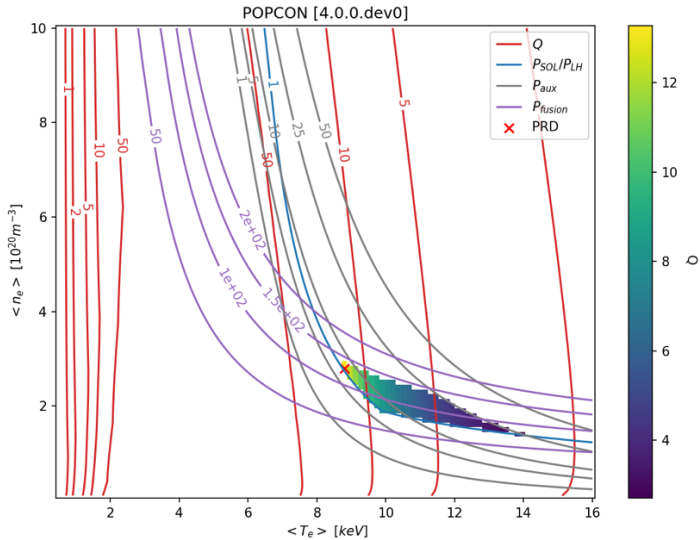

*Figure 2: Updated plasma operating contour (POPCON) for full-field, full-current H-mode operation in SPARC. Red contours are Q, purple contours are fusion power in MW, the grey contours are the required auxiliary heating power in MW and the blue contour is the L–H threshold power. The filled region represents the operational space where SPARC is above the L–H power threshold but below the available auxiliary heating power and below a $P_{fusion} < 140MW$ limit to prevent excessive heating of the magnets. The colormap indicates the fusion power gain across the operational space. Temperature and density are the volume-averaged values. The red cross is the operating point for the full-performance H-mode discharge.*

## Comparing the two design points.

SPARC Physics Basis values are from Table 2 from Creely, JPP, 2020 for the "Full-field H-mode".

POPCON Release values are for a 100x100 grid from $\langle T_e \rangle \sim 0.5 - 16 keV, \langle n_e \rangle \sim 0.1 - 10 \times 10^{20} m^{-3}$.

|  | SPARC Physics Basis | POPCON Release |  | Notes |
|---|---|---|---|---|
| $B_0$ | 12.2 | 12.2 | $T$ |  |
| $I_p$ | 8.7 | 8.7 | $MA$ |  |
| $q^*_{Uckan}$ | 3.05 |  |  |  |
| $q^*_{ITER}$ |  | 3.29 |  |  |
| $\rho^*$ | 0.0027 | 0.0029 |  |  |
| $\nu_{eff}$ | 0.16 | 0.098 |  |  |
| $\nu^*$ | 0.029 | 0.019 |  |  |

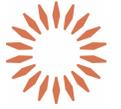

|  |  |  |  |  |
|---|---|---|---|---|
| $H_{98y,2}$ | 1.0 | 1.0 |  |  |
| $\tau_e$ | 0.77 | 0.65 | $s$ |  |
| $P_{RF}$ | 11.1 | 7.22 | $MW$ |  |
| $P_{Ohmic}$ | 1.7 | 1.28 | $MW$ |  |
| $Z_{eff}$ | 1.5 | 1.34 |  | Now computed from impurities |
| $n_{DT}/n_e$ | 0.85 | 0.85 |  | Now computed from impurities |
| $\langle T_e \rangle$ | 7.3 | 8.63 | $keV$ |  |
| $\langle T_i \rangle$ | 7.3 | 8.63 | $keV$ |  |
| $\langle n_e \rangle$ | 3.1 | 2.9 | $10^{20} m^{-3}$ |  |
| $\langle n_i \rangle$ | 2.7 | 2.5 | $10^{20} m^{-3}$ |  |
| $\nu_{T_e}$ | 2.5 | 2.5 |  |  |
| $\nu_{T_i}$ | 1.33 | 1.36 |  |  |
| $f_G$ | 0.37 | 0.34 |  |  |
| $\beta$ | 0.012 | 0.013 |  |  |
| $\beta_N$ | 1.0 | 1.0 | $m \cdot T/MA$ |  |
| $P_{sep} B_0 / R_0$ | 191 | 184 | $MW \cdot T/m$ |  |
| $P_{fusion}$ | 140 | 136 | $MW$ |  |
| $Q$ | 11.0 | 14.35 |  |  |

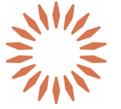

Comparing inputs from SPARC Physics Basis to cfsPOPCON v4.0.0 release.

| Input | Value | | Unit |
|---|---|---|---|
| $R_0$ | 1.85 | 1.85 | $m$ |
| $B \cdot R$ | 22.5 | 22.5 | $T \cdot m$ |
| $\epsilon$ | 0.3081 | 0.3081 | |
| $\kappa_A$ | 1.75 | 1.75 | |
| $\delta_{95}$ | 0.3 | 0.3 | |
| $I_p$ | 8.7 | 8.7 | $MA$ |
| $a/L_{T_e}$ | 2.5 | 2.5 | |
| $\nu_T$ | 2.5 | 2.5 | |
| $n_{DT}/n_e$ | 0.85 | 0.85 (computed) | |
| $Z_{eff}$ | 1.5 | 1.34 (computed) | |
| $n_D/(n_D + n_T)$ | 0.5 | 0.5 | |
| $c_W$ | 1.5e-5 | 1.5e-5 | |
| $c_{He}$ | 6e-2 | 6e-2 | |
| $c_O$ | 3.1e-3 | 3.1e-3 | |
| $H_{98y2}$ | 1.0 | 1.0 | |
| $P_{rad}$ method | $2.25 \times P_{Brems.}$ | $P_{Brems.} + P_{Synch.} + P_{Line}$ | |
| $\nu_{n_e}$ offset | -0.1 | -0.1 | |
| $\nu_{n_i}$ offset | -0.2 | -0.2 | |
| $P$ in $\tau_e$ calc. | $P_{SOL}$ | $P_{in}$ | |
| Form of profiles | Parabolic | Scaled from TRANSP | |

There are three key differences in the inputs used to generate these design points — namely, using $P_{in}$ instead of $P_{SOL}$ in the $\tau_e$ computation, using scaled-TRANSP profiles instead of parabolic profiles, and computing the impurity radiated power directly instead of increasing the Bremsstrahlung losses. To study the effect of these changes, we selected a version of cfsPOPCON from the git history (not publicly available) which reproduces the result in Figure 1.

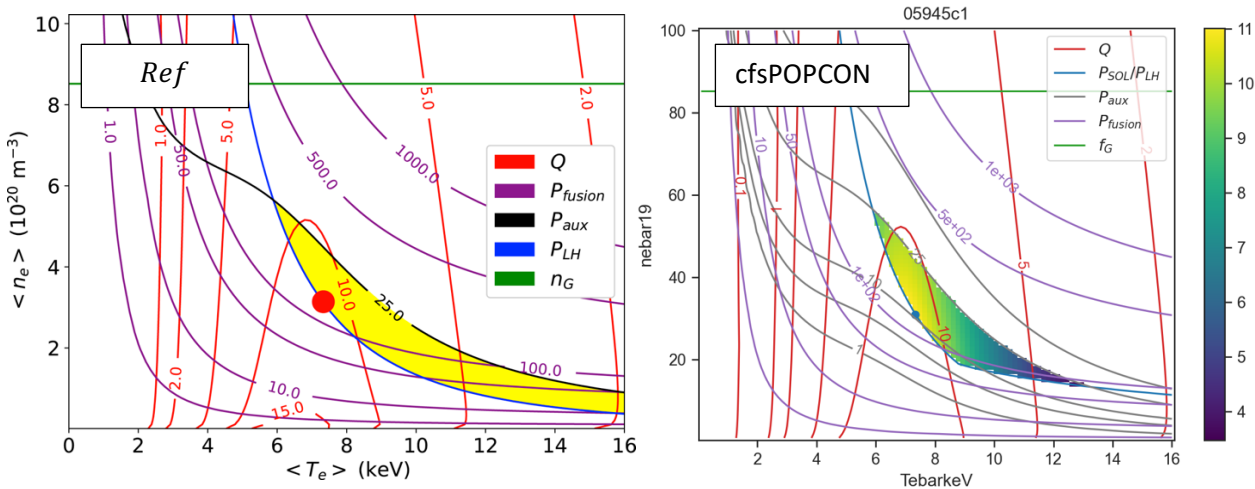

The two results are identical except that there is now no H-mode operational space below $\langle n_e \rangle \sim 1.5 \times 10^{20} m^{-3}$. This is due to a low-density branch [3] switch implemented in the calculation of $P_{LH}$. The following section is written using this version of cfsPOPCON.

## Using $P_{in}$ instead of $P_{SOL}$ in the $\tau_e$ computation.

In Chapter 2 of the ITER Physics Basis where the ITER98y,2 energy confinement scaling is introduced, it is stated that "*With respect to other factors that may influence the confinement time, but which are poorly accounted for in the dataset, we mention that, for practical reasons, the power lost by radiation inside the separatrix of the existing devices has been neglected when deriving the scalings*" [2].

For this reason, we decided to switch from using $P_{SOL} = P_{in} - P_{rad,core}$ to using $P_{in}$ in the calculation of $\tau_e$. To understand the impact of this, we need to introduce how $\tau_e$ and $P_{in}$ (or $P_{SOL}$) are computed from $W_p$ in cfsPOPCON. The cfsPOPCON algorithm may be considered "back-to-front", in that it starts by selecting an average density and average temperature pair and then computing the power that would be needed to sustain that operational point.

We can write two separate expressions for $\tau_e$, in terms of a $\tau_e$ scaling or in terms of the plasma stored energy. Defining some arbitrary constant $C_1$ (fixed for a given $\langle n_e \rangle$ and machine parameters such as $I_p$) we can write the energy confinement time $\tau_e$ in terms of a power $P$, an exponent $\alpha$ ($= -0.69$ for ITER98y,2) and the plasma stored energy $W_p \sim \langle n_e \rangle \langle T_e \rangle$:

$$\tau_e = C_1 P^\alpha = W_p/P$$
$$P = \left(\frac{W_p}{C_1}\right)^{1/(\alpha+1)} = C_2 W_p^{3.2}$$

For a given point on a POPCON with a given $\langle n_e \rangle, \langle T_e \rangle$ pair, that power $P$ is fixed regardless of whether we decide it should be $P_{in}$ or $P_{SOL}$.

If we assume that $P = P_{SOL}$, we need $P_{rad} = P_{in} - P_{SOL}$ *more power* to stay at any given point — compared to if we assume that $P = P_{in}$.

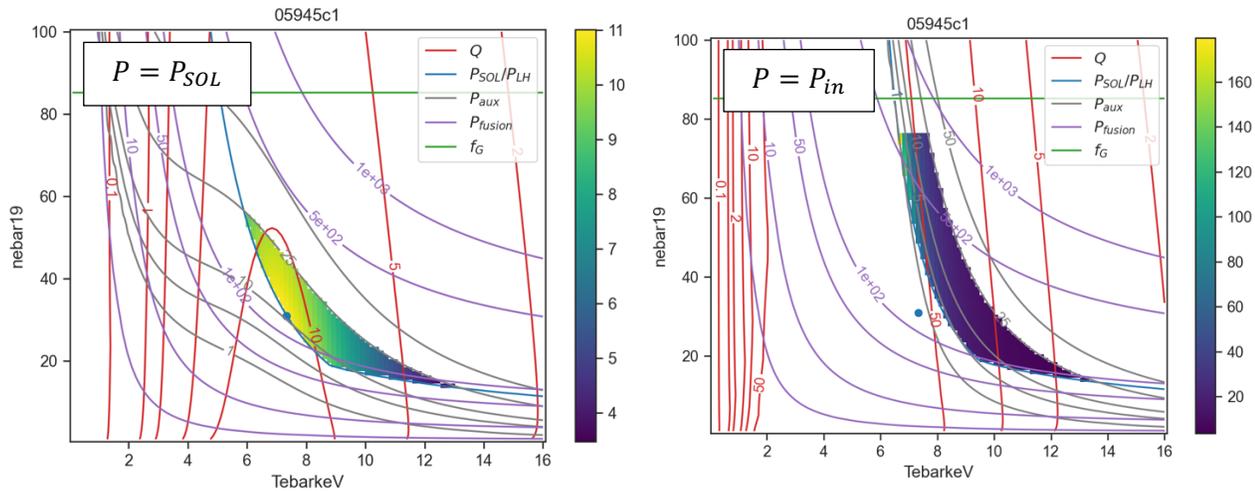

In the two above plots, we show the impact of switching this assumption. Clearly, this change in assumption has an enormous impact. Our fusion gain $Q_{absorbed} = P_{fusion}/P_{external}$ has increased dramatically, since we require less power to maintain a given $P_{fusion} = f(\langle n_e \rangle, \langle T_e \rangle)$ and therefore have reduced our denominator while keeping the numerator the same. Additionally, we can push to higher densities since there's much more space with $P_{RF} < P_{RF,max} = 25 MW$.

Finally, our entire operational space is shifted to higher temperatures because we now have less $P_{SOL}$ for a given $\langle n_e \rangle, \langle T_e \rangle$ point and we need $P_{SOL} > P_{LH}$ to stay in H-mode. It's worth thinking about this a little more: at a point where we would previous have had $P_{SOL} = P$ (where $P$ is our undefined power from the $\tau_e$ scaling), we now have $P_{SOL} = P - P_{rad}$. Therefore, to achieve $P_{SOL} > P_{LH}$ we need to go to higher $P = C_2 W_p^{3.2}$ — which explains the shift in the operational space.

The SPARC Primary Reference Discharge defined by cfsPOPCON                                                                7

## Switching from parabolic to scaled TRANSP H-mode-like profiles.

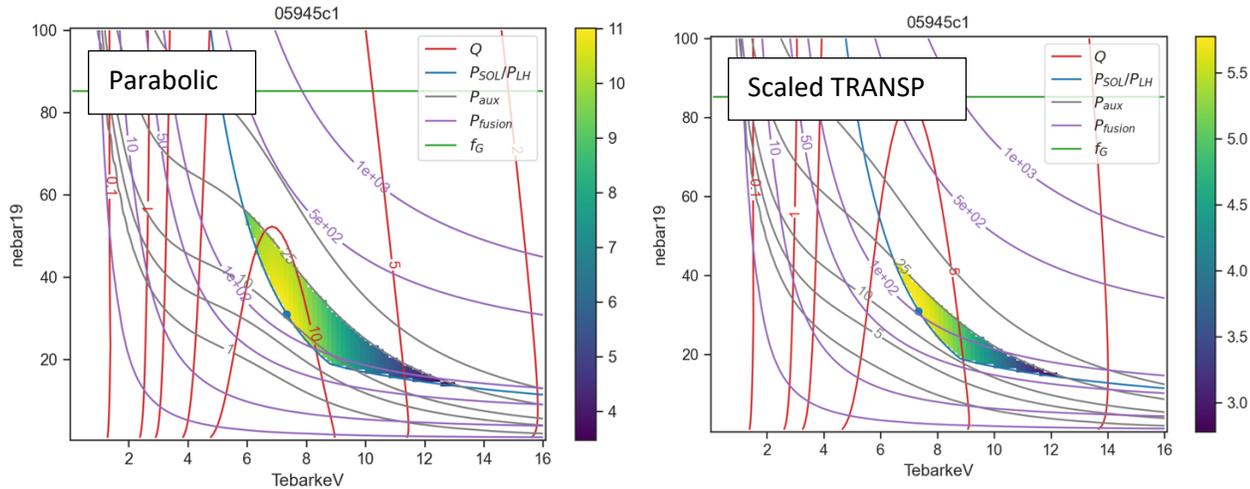

Compared to realistic H-mode profiles, parabolic profiles have a larger volume-weighted peaking for the same core-to-volume-average peaking factor. Because of this, they overestimate the fusion power. To correct for this, we switched to scaling numerical results from TRANSP studies of SPARC at PRD parameters [4]. Unsurprisingly, this reduced the maximum achievable $Q$ due to the reduced fusion power at a given $\langle n_e \rangle$, $\langle T_e \rangle$ point.

## Including the line- and synchrotron-radiated power contributions

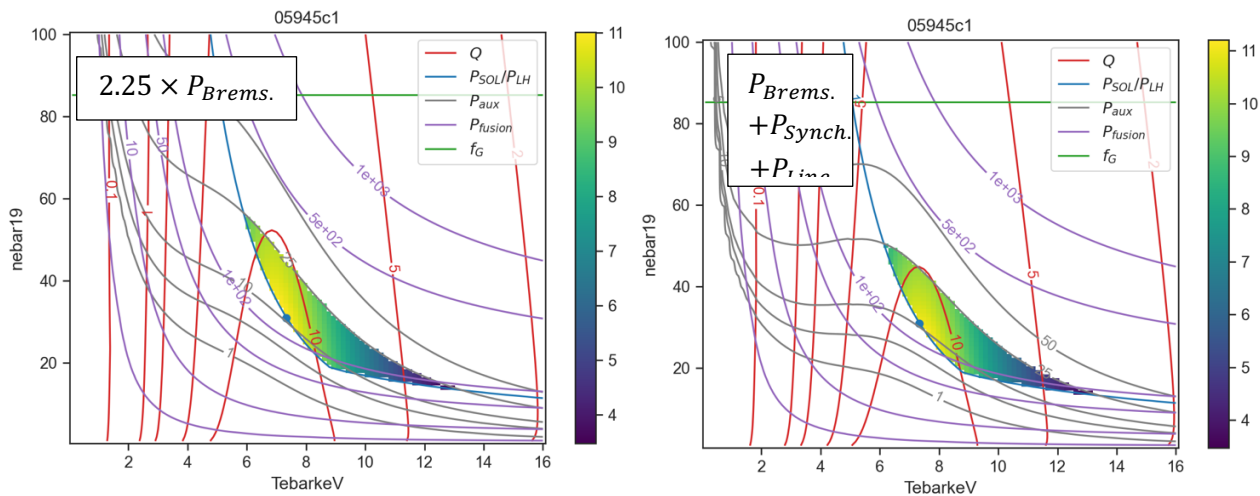

The original PRD design point was defined assuming that power radiated from the confined region was equal to the Bremsstrahlung power (for the given $Z_{eff} = 1.5$), scaled up by a factor of 2.25. This has been replaced by consistently including the line radiation from impurities as well as synchrotron losses. This change does not have much of an affect the highest performance point, but it does change the shape of the low-temperature contours.

## The combined effect of changing the input assumptions.

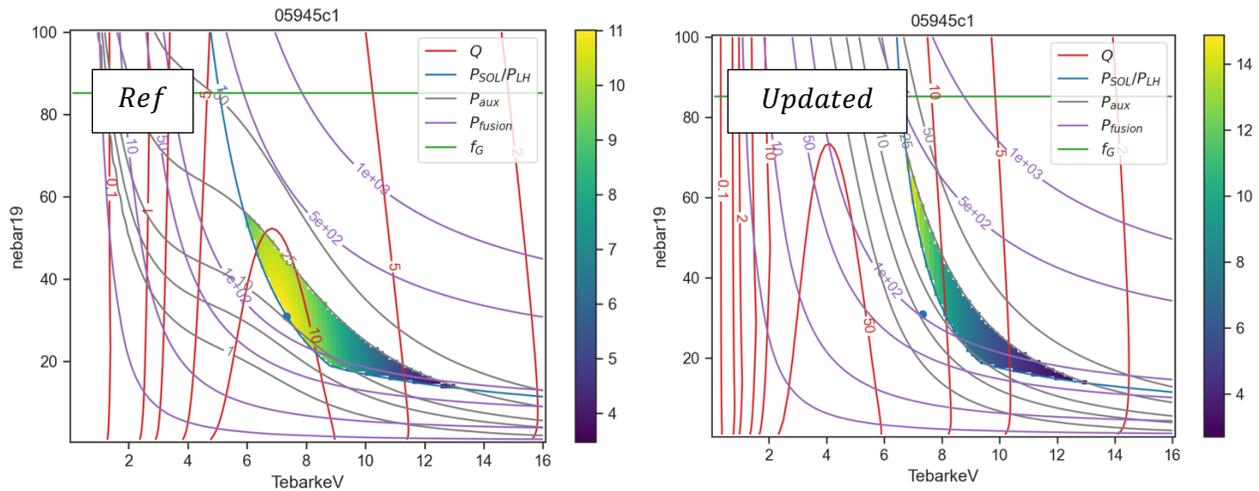

The increased performance (maximum achievable $Q$) from switching from $P_{SOL}$ to $P_{in}$ in the $\tau_e$ computation is largely counteracted by the drop in performance from switching to realistic H-mode profiles. If we further eliminate the high-density regions with $P_{fusion} > 140 MW$ (too much neutron heating of magnets), the operational space looks very similar even though we've significantly changed our assumptions.

## Smaller changes in algorithms

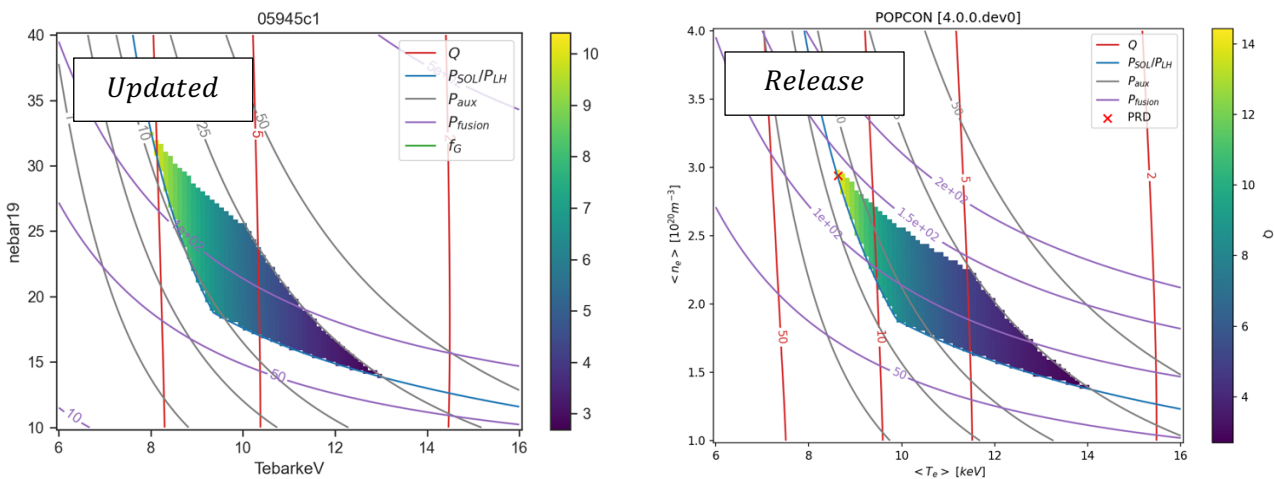

Changing these three input assumptions accounts for the largest changes to the PRD (shown as "Updated", matching the right figure above except now focusing on the accessible operational space). There are a series of smaller improvements which were made since then which slightly change the result obtained by the released v4.0.0 version of cfsPOPCON (shown as "Release").

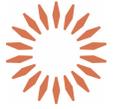

The differences between these two versions are:

1. Minor corrections to unit-conversion constants,
2. Fixed an error in the calculation of $\beta_{poloidal}$,
3. Using inductive rather than total current to calculate the ohmic heating power,
4. Using $\beta_{tot}$ instead of $\beta_e$ in the calculation of the density peaking,
5. Consistently calculating $Z_{eff}$ and $n_{DT}/n_e$ from the specified impurities,
6. Changing from calculating $Q_{absorbed}$ to $Q_{launched}$ by adding a penalty factor on the auxiliary power. Since we're primarily RF-heated, we use $f_{coupled} = 0.9$ which corresponds to the power coupling found for 6% $^3He$ in the SPARC Physics Basis [5]. This reflects a decision by CFS to only use the launched power when computing the fusion power gain.